\documentclass[pra,twocolumn,showpacs,amsmath,amssymb,superscriptaddress]{revtex4-1}

\usepackage{graphicx}
\usepackage{dcolumn}
\usepackage{bm}

\usepackage{amsmath}	
\begin{document}
\draft
\newcommand{\lw}[1]{\smash{\lower2.ex\hbox{#1}}}

\title{Phase-separated ferromagnetism in spin-imbalanced Fermi atoms 
loaded on an optical ladder: a density matrix renormalization group
study} 

\author{M.~Okumura}
\email{okumura@riken.jp}
\affiliation{Computational Condensed Matter Physics Laboratory, RIEN,
Wako, Saitama 351--0198, Japan} 
\affiliation{CREST(JST), 4--1--8 Honcho, Kawaguchi, Saitama 332--0012,
Japan}
\author{S.~Yamada} 
\affiliation{CCSE, Japan Atomic Energy Agency, Higashi-Ueno, 
Tokyo 110--0015, Japan}
\affiliation{CREST(JST), 4--1--8 Honcho, Kawaguchi, Saitama 332--0012,
Japan}
\author{M.~Machida}
\affiliation{CCSE, Japan Atomic Energy Agency, Higashi-Ueno, 
Tokyo 110--0015, Japan}
\affiliation{CREST(JST), 4--1--8 Honcho, Kawaguchi, Saitama 332--0012,
Japan}
\author{H.~Aoki}
\affiliation{Department of Physics, University of Tokyo, Hongo, Tokyo
113--0033, Japan}

\date{\today}

\begin{abstract} 
 
 We consider repulsively-interacting cold fermionic atoms loaded on an
 optical ladder lattice in a trapping  potential. The density-matrix 
 renormalization-group method is used to numerically calculate the
 ground state for systematically varied values of interaction $U$ and  
 spin imbalance $p$ in the Hubbard model on the ladder. The system
 exhibits rich structures, where a fully spin-polarized phase,
 spatially separated from other domains in the trapping potential,
 appears for large enough $U$ and $p$. The phase-separated
 ferromagnetism can be captured as a real-space image of the energy
 gap between the ferromagnetic and other states arising from a
 combined effect of Nagaoka's ferromagnetism extended to the ladder
 and the density dependence of the energy separation between competing
 states. We also predict how to maximize the ferromagnetic region.
\end{abstract}

\maketitle
{\it Why cold atoms on a ladder? ---} 
Ultracold atom systems offer ideal opportunities for systematic
studies of novel quantum many-body phenomena, since they are not only
extremely clean, but also controllable to an unusual degree
\cite{CAreview}. Particularly, the inter-atomic interaction tunable
with the Feshbach resonance create a unique stage for exploring  
strongly-correlated systems. If we further turn to cold atoms on
optical lattices (OL's) prepared by standing waves from laser beams,
they provide an even more versatile playing ground
\cite{CAreview,OLreview}. Among the attractive targets are the
metal-insulator transition, $d$-wave superfluidity, and various
magnetisms.  Actually, the antiferromagnetism via the superexchange
interactions between localized Bose atoms on an optical superlattice
\cite{SuperExExp} and the Mott insulator in fermionic atoms on cubic
OL \cite{MottCoreExp1,MottCoreExp2} have already been achieved.

Now, one big issue in the field of strongly correlated systems is the
itinerant ferromagnetism, which is, despite the long history, still
far from fully understood. Cold atom systems, with their tunability,
should be an unprecedented place for realizing the itinerant
ferromagnetism. While there are many theoretical studies on itinerant
magnetism in ultracold atom systems
\cite{MagSB1,MagSB2,Conduit,LeBlanc,Wunsch,Stecher}, Jo {\it et al.}
recently reported that the Stoner instability was observed in an
ultracold fermionic atom system without OL \cite{StonerExp}. However,
it has been pointed out that some factors other than Stoner
instability may also be involved \cite{Pekker}. 

\begin{figure}
\includegraphics[scale=0.7]{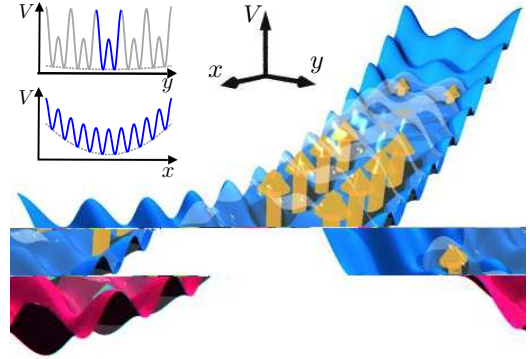}
\caption{\label{fig1} (Color online) Two-leg ladder optical lattice
 (corrugated surface in blue) is schematically shown along with the
 atomic wavefunction (pale cloud) and spins (arrows). Inset depicts
 simple lattice (along $x$) and superlattice ($y$) potentials for
 creating the optical ladder.} 
\end{figure}

In this Rapid Communication, we study strongly-correlated fermionic
atoms with a spin-imbalance loaded on an optical ladder, and propose a
way to realize and observe the ferromagnetism originated by the finite
hole-density Nagaoka mechanism.
One tunability unique to cold atoms is we can control the spin
balance, which enriches the quantum states. For example, a phase
separation between superfluid and normal phases was observed in
attractively interacting ultracold Fermi atoms
\cite{SIBMIT,SIBRice}. On the other hand, a magnetic structure with
spin imbalance in repulsively interacting fermionic atoms with and
without OLs has been studied in a weak-correlation regime 
\cite{Conduit,LeBlanc,Wunsch}.  

\begin{figure*}
\includegraphics[scale=1]{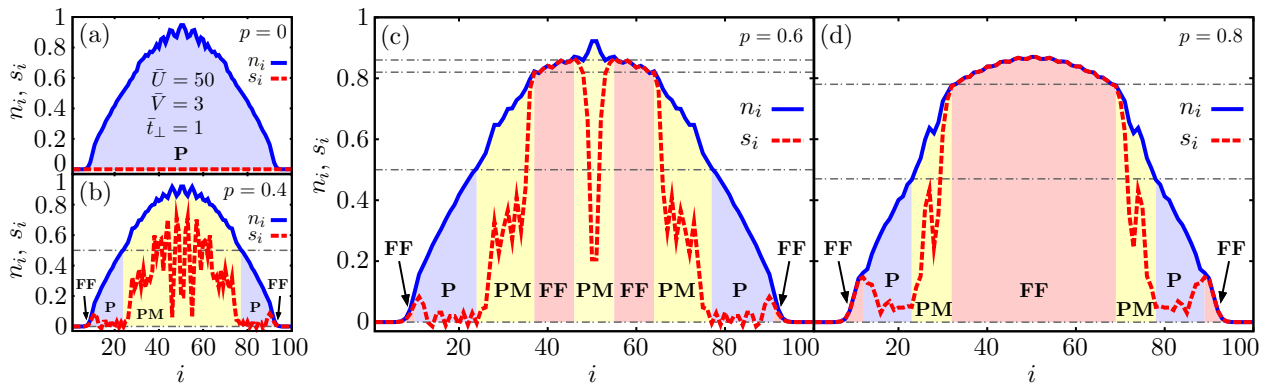}
\caption{\label{fig2} (Color online) The particle density ($n_i$; blue
 solid line) and spin density ($s_i$; red dashed line) for the spin
 polarization $p=0$ (a) , $0.4$ (b) , $0.6$ (c), and $0.8$ (d) with
 $\bar{U}=50$, $\bar{V}=3$, $\bar{t}_\perp = 1$, and $L=100$. P, PM,
 and FF denote paramagnetic, partially magnetic, and fully
 ferromagnetic states, respectively. The dash-dotted lines represent
 the $n_i$ at numerically obtained phase boundaries.} 
\end{figure*}

The reason why we take the ladder is as follows.
The long history of itinerant ferromagnetism has made us
realize that Stoner's picture is mean-field theoretic, while almost
the only model in which itinerant ferromagnetism can be rigorously
shown is Nagaoka's ferromagnetism, which requires rather a
pathological (infinitely strong interaction and an infinitesimal 
doping in a half-filled band) limit \cite{Nagaoka}. The Nagaoka 
ferromagnetism has yet to be experimentally confirmed, although von
Stecher {\it et al.} recently proposes a way to realize the Nagaoka
ferromagnetism in optical plaquette systems with the high tunability
of OL system \cite{Stecher}. On the other hand, some theories indicate
that the two-leg ladder can realize an itinerant ferromagnetism that 
accommodates finite interaction and finite doping, which is connected
to Nagaoka's in the limit of $U \rightarrow \infty$ or large inter-leg
transfer \cite{Kohno,Arita}. Experimentally, an optical ladder may be
created by a superposition of normal lattice potential (along $x$) and
super-lattice potential ($y$) (Fig.~\ref{fig1}). This should be
realizable because a superlattice potential has been used in the
experiments with bosonic atoms \cite{SL,SuperExExp}. To confine the
atoms we impose a trapping potential as well, and we explore the 
ground states in this situation. We want to treat very strong
repulsive interactions, but the system, being quasi-one dimensional, 
can be treated with the density-matrix renormalization-group (DMRG)
method \cite{White,DMRGreview}, which can also treat effects of 
the trapping potential \cite{dexDMRG}. We show that an itinerant
ferromagnetism does appear in an optical ladder with an interesting
phase separation between a fully-polarized ferromagnetic domain
originating from the extended Nagaoka mechanism with finite hole
density. This ground state should be easily observed by {\it in situ}
imaging method \cite{SIBMIT}, which is the main reason why we consider
the spin-imbalance. In addition, we find the enhancement of the
formation of the finite hole-density Nagaoka ferromagnetism by the
spin-imbalance.

{\it Formulation ---} 
We take the simplest possible repulsive Hubbard model on a ladder in a
trapping potential applied along the leg. The Hamiltonian then reads,
in the standard notation, 
\begin{align}
 \hat{\cal H} & = -t \sum_{\langle i,j \rangle, \alpha, \sigma}
 \hat{c}^{\alpha\dag}_{i\sigma} \hat{c}_{j\sigma}^{\alpha \phantom
 \dag} - t_\perp \sum_{i, \langle \alpha, \beta \rangle, \sigma}
 \hat{c}^{\alpha\dag}_{i\sigma} \hat{c}_{i\sigma}^{\beta \phantom
 \dag} \nonumber \\  
 & \quad {} + U \sum_{i,\alpha} \hat{n}_{i\uparrow}^\alpha
 \hat{n}_{i\downarrow}^\alpha + V \sum_{i, \alpha, \sigma} \left( i - x_{\rm
 c} \right)^2 \hat{n}_{i\sigma}^\alpha , \label{Hamiltonian}   
\end{align}
where, $i,j=1,2,\cdots,L$ ($L$: the ladder length) label the sites
along the legs with $\langle i,j \rangle$ being nearest-neighbor sites
connected by a hopping integral $t$, $\alpha=1,2$ labels the sites
along the rungs with $t_\perp$ being the hopping along the
rung.  $U(\ge 0)$ is the repulsive, on-site interaction \cite{Unote}, 
while $V$ is the strength of the trapping potential, which is assumed
to be harmonic around the center, $x_{\rm c} = (L+1)/2$ with the
lattice constant taken to be unity. It should be mentioned that
the maximum value of $\bar{U}$ for stable Hubbard model in the current
experiment is 180 \cite{MottCoreExp1}. We assume that the harmonic
confinement along the rung (superlattice) direction is very weak. In
addition, we also assume that the inter-ladder hoppings are
negligible.

The total spin imbalance is defined as $p = (N_\uparrow -
N_\downarrow)/N$, where $N_\sigma = \sum_{i,\alpha}
\langle \hat{n}_{i\sigma}^\alpha \rangle$ and $N = \sum_\sigma
N_\sigma$, and $\langle \cdot \rangle$ represents the
expectation value. All the numerical results respect the symmetry,
$\langle \hat{n}_{i\sigma}^1 \rangle = \langle \hat{n}_{i\sigma}^2
\rangle$, so that we can introduce the rung particle density $n_i =
\langle \hat{n}_{i\uparrow}^1 \rangle + \langle
\hat{n}_{i\downarrow}^1 \rangle$, and the rung spin density $s_i
\equiv \langle \hat{n}_{i\uparrow}^1 \rangle - \langle
\hat{n}_{i\downarrow}^1 \rangle$. For convenience, we use
dimensionless parameters as $\bar{t}_\perp = t_\perp / t$,
$\bar{U}\equiv U/t$, and $\bar{V} \equiv V/t \times 10^3$. In this
paper, we fix the total number of atoms as $N=100$ with the length of
the ladder $80 \leq L \leq 180$. The number, $m$, of the states
retained in the present DMRG calculations is $800$--$1200$, which are
numerically shown to give converged results.  

{\it DMRG results ---} 
Let us show the DMRG results. Figure \ref{fig2} shows profiles of
$n_i$ and $s_i$ in the ground states for various $p$ in a
strongly-interacting case ($\bar{U} = 50$) with $\bar{t}_\perp = 1$
and $\bar{V}=3$. In the spin-balanced case ($p = 0$)
[Fig.\ref{fig2}(a)], a paramagnetic (P) phase appears. As we increase
$p$ to $0.4$, a partially magnetic (PM) phase starts to appear in the
middle of the system [Fig.~\ref{fig2}(b)]. If we further increase $p$
to $0.6$, fully ferromagnetic (FF) regions emerge in the ground state
[Fig.~\ref{fig2}(c)] in a more complex phase-separated structure. For
$p = 0.8$, the FF region covers almost the whole system, sandwiched by
narrow PM and P regions [Fig.~\ref{fig2}(d)]. 

\begin{figure}
\includegraphics[scale=1]{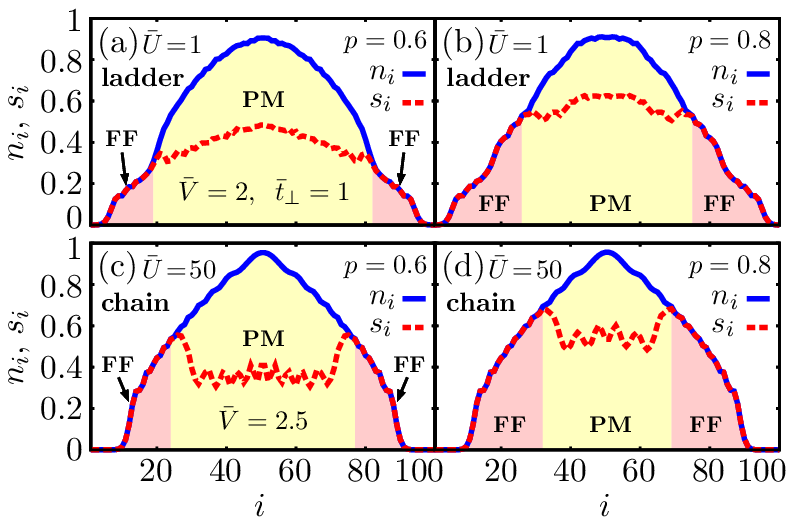}
\caption{\label{fig3} (Color online) The particle density $n_i$ (blue
 solid lines) and spin density $s_i$ (red dashed lines) for a 
 weakly-interacting ladder system ($\bar{U}=1$, $\bar{V}=2$,
 $\bar{t}_\perp = 1$, and $L=100$) with $p = 0.6$ (a) or $0.8$ (b),
 and for a strongly-interacting single chain ($\bar{U}=50$ and
 $\bar{V}=2.5$)  with $p = 0.6$ (c) or $0.8$ (d) for $N = 50$.} 
\end{figure}

In order to demonstrate both of the effect of strong interaction and
the effect of the ladder configuration (a connectivity condition
\cite{Nagaoka}) are at work for realization of the FF phases in the
regions with $0.8 < n < 0.9$, we compare the result with the one 
in a weakly-interacting case with connectivity [two-leg ladder,
$\bar{U} = 1$, $\bar{V} = 2$, and $\bar{t}_\perp = 1$;
Figs.~\ref{fig3}(a)(b)], and with the one in a strong-interacting case 
without connectivity [single chain, $\bar{U}=50$, and $\bar{V}= 2.5$;
Figs.~\ref{fig3}(c)(d)]. We can immediately see that FF phases do not
emerge in these systems, although there are narrow FF regions around
the edges. This clearly indicates that the FF phases with $0.8 < n
<0.9$ in the ladder arise due to a combined effect of the strong
interaction and the connectivity condition, i.e., the finite
hole-density Nagaoka ferromagnetism \cite{Kohno}. 

\begin{figure}
\includegraphics[scale=1]{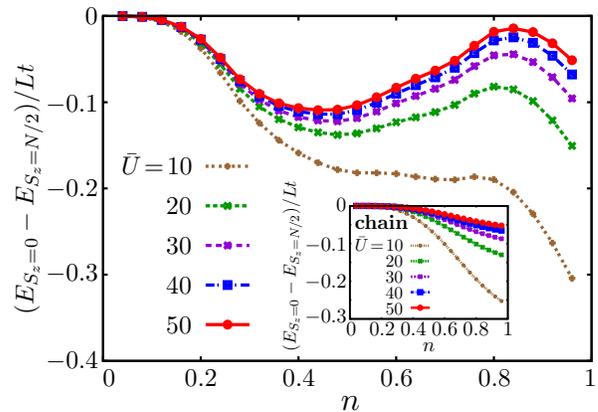}
\caption{\label{fig4} (Color online) The density dependence of the
 energy gap $\Delta \bar{E} = (E_{S_z=0}-E_{S_z=N/2})/Lt$ in
 homogeneous systems ($\bar{V} = 0$) with  $\bar{t}_\perp = 1$ for
 various values of $U$. The inset is the same plot for the homogeneous
 single chain. Here the parameters are $L=50$, $\bar{t}_\perp = 1$ and 
 $m=1200$ (ladder) or $m=1000$ (chain).} 
\end{figure}

Now, the question is what determines the spatial distribution of 
majority and minority spins as the spin imbalance is increased. As a
key, we can examine the energy gap, $\Delta \bar{E} = (E_{S_z = 0} -
E_{S_z = N/2}) / L t$, of the FF above the lowest energy of the
nonmagnetic state in the uniform system ($V = 0$), as in the $t$--$J$
model in \cite{Kohno} but here for the Hubbard model. The gap depends
on the particle density $n$ as depicted in Fig.~\ref{fig4}. As $U$ is
increased the curve approaches the $\Delta \bar{E} = 0$ axis around $n
\sim 0.8$, which means that the ferromagnetic phase becomes closer to
the ground state. Such a non-trivial structure is naturally absent in
the single-chain Hubbard model, as shown in the inset of the figure. 

With this picture we can interpret the results shown in
Figs.~\ref{fig2}, \ref{fig3}(c) and \ref{fig3}(d). When the trapping  
potential is not too strong so that the spatial variation of the
density is slow, we can locally define the energy gap as given in
Fig.\ref{fig4}. The energy gap and the trapping potential determine
the phase separation: Let us first look at the strongly interacting 
1D chain, both of the majority ($=(n_i+s_i)/2$) and minority
($=(n_i-s_i)/2$) spins are accumulated in the middle where $n 
\simeq 1$ [Figs.~\ref{fig3}(c)(d)]. Non-FF states are preferred
because both of $\Delta \bar{E} (n)$ (Fig.~\ref{fig4}, inset) and the
trapping potential favor $n \simeq 1$, which is a realization of the
Lieb--Mattis theorem \cite{LiebMattis}. This leads us to a natural
interpretation that the ground-state energy is lowered by gathering
minority spins around the trap center to construct an
antiferromagnetic correlation. On the other hand, in a
strongly-interacting ladder, we obtain the ferromagnetic region, which 
may seem counterintuitive, but is in fact viewed as a manifestation of 
the extended Nagaoka mechanism \cite{Kohno}. Namely, the minority
spins in Fig.~\ref{fig2}(d) reside in the regions with relatively high
potential energy ($i < 31$ and $70 < i$), but this is compensated by a
large energy $\Delta \bar{E}$ gained by forming the non-FF state
(Fig.~\ref{fig4}). Conversely, FF state is stabilized for $0.8 < n <
0.9$, where $\Delta \bar{E}$ is small. Hence the phase separation
arises. By contrast, the phase separation seen for a weak $U$ in the
ladder in Figs.~\ref{fig3} (a)(b) has a simple origin, i.e., both
majorities and minorities are accumulated by the trapping potential
and form doubly occupied states because the inter-particle interaction
is weak. Finally, the spin-imbalance enhances the formation of the
finite hole-density Nagaoka ferromagnetism, i.e., the FF regions with
$0.8 < n < 0.9$ appear when $p = 0.6$ and $0.8$
[Figs.~\ref{fig2}(c)(d)], although they disappear in the spin-balanced
case [Fig.~\ref{fig2}(a)] due to a small gap $\Delta \bar{E}$ around
$n \sim 0.8$ (Fig.~\ref{fig4}) \cite{Okumura}.

\begin{figure}
\includegraphics[scale=1]{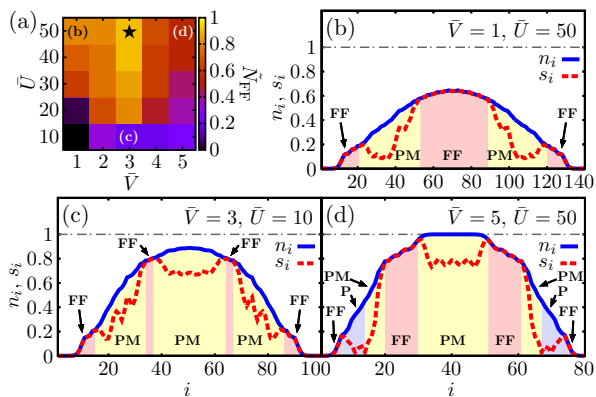}
\caption{\label{fig5} (Color online) (a) $\tilde{N}_{\rm FF}$ plotted
 against $\bar{V}$ and $\bar{U}$ for $\bar{t}_\perp = 1$ and $p =
 0.8$. In the color-coding $\tilde{N}_{\rm FF}$ is normalized by the
 maximum [a star at $\bar{V} = 3, \bar{U} = 50$, which corresponds to 
 Fig.~\ref{fig2}(d)] in the parameter region considered. Also plotted
 are $n_i$ (blue solid lines) and $s_i$ (red dashed) with $(\bar{V},
 \bar{U}) = (1, 50)$ (b), $(5, 50)$ (c), and $(3, 10)$ (d).} 
\end{figure}

{\it Phase diagram ---} 
Having clarified the mechanism for the phase-separated magnetism, we
now explore a phase diagram against $\bar{U}$, $\bar{V}$, and
$\bar{t}_\perp$. First, we focus on the dependence on $\bar{V}$ and
$\bar{U}$ for  $p = 0.8$ with $\bar{t}_\perp = 1$  in
Figs.~\ref{fig5}(a). We characterize the FF region by $\tilde{N}_{\rm
FF}$, which  is the number of sites having $n_i = s_i$ within
$10^{-2}$ normalized by the maximum value of this quantity. We find
that there is optimum $\bar{U} \simeq 50$ and $\bar{V} \simeq 3$ to
make $\tilde{N}_{\rm FF}$ maximum. To trace back how the optimum
$\bar{U}$ and $\bar{V}$ arise, we can look at $n_i$ and $s_i$ plotted
for various values of $\bar{V}$ and $\bar{U}$ in
Figs.~\ref{fig5}(b)-(d). If we first combine Figs. \ref{fig2}(d),
\ref{fig5}(b)(d) for the effect of varied $\bar{V}$ with a fixed
$\bar{U} = 50$, we can see that the central FF region broadens when
$\bar{V}$ increases form $1$ to $3$
[Figs.~\ref{fig2}(d),~\ref{fig5}(b)] because the region with $n_i >
0.8$ becomes wider with $\bar{V}$. If the trap becomes too strong in
Fig.~\ref{fig5}(d), however, the FF region gives way to the PM Mott
plateau because $n_i$ reaches 1. If we turn to dependence on the
interaction strength, we can look at the results for $\bar{U} =
10$--$50$ with a fixed $\bar{V} = 3$ [Figs. \ref{fig2}(d),
\ref{fig5}(c)].  As the interaction is decreased to 10 in
Fig.~\ref{fig5}(c) the FF region shrinks, where the reason should be
that $\Delta \bar{E}$ becomes large at $n \sim 0.8$ when $\bar{U}$ is
as small as $10$. 

\begin{figure}
\includegraphics[scale=1]{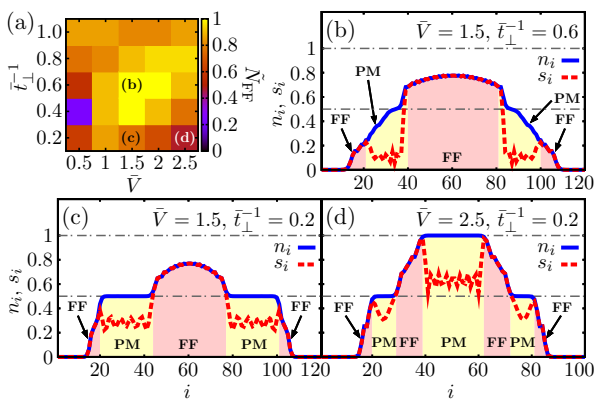}
\caption{\label{fig6} (Color online) (a) Color-coded $\tilde{N}_{\rm
 FF}$ against $\bar{V}$ and $\bar{t}_\perp^{-1}$ for $\bar{U} = 50$
 and $p = 0.8$. Also plotted are $n_i$ (blue solid lines) and $s_i$
 (red dashed) with $(\bar{V}, \bar{t}_\perp^{-1}) = (1.5, 0.6)$ (b),
 $(1.5, 0.2)$ (c), $(2.5, 0.2)$ (d).} 
\end{figure}

Finally we consider the effect of $\bar{t}_\perp$. A phase diagram in
terms of $\tilde{N}_{\rm FF}$, against  $\bar{V}$ and
$\bar{t}_\perp^{-1}$ this time, is displayed in Fig.~\ref{fig6}(a). We
find that a finite $\bar{t}_\perp$ is suitable for largest
$\tilde{N}_{\rm FF}$ in this figure, which is contrast to the
homogeneous system in which the infinite $\bar{t}_\perp$ is best. The
reason is that a charge gap opens at $n_i = 0.5$ \cite{Kohno}, at
which the FF phase gives way to the insulating state when
$\bar{t}_\perp^{-1}$ is small [Figs.~\ref{fig6}(b)--(d)]. This
degrades the FF phase stability.  

{\it Summary ---} 
We have found for spin-imbalanced and strongly repulsive-interacting
Fermi atoms on an optical ladder that ferromagnetic regions appear in
a phase-separated structures. Their emergence is caused by a combined
effect of the strong interaction and the connectivity condition, i.e.,
the extended Nagaoka mechanism. The origin of the phase separation is
explained by the density dependence of the energy gap between fully
spin-polarized and other states in homogeneous system. As for finite
temperatures, it was shown, with the dynamical mean-field theory, that
the finite hole-density Nagaoka ferromagnetism is robust against
thermal fluctuations in two-dimensional system \cite{DMFT}. Then we
expect the phase separation is also robust at finite temperatures. 
Another future problem is how the itinerant ferromagnetism treated here 
in the spin-imbalanced condition would be related to the balanced
case. Along the line, a study of correlation functions in
spin-balanced case is under way \cite{Okumura}, where we observe phase
separation between non-magnetic and ferromagnetic (for the $S_x$ and
$S_y$ direction) phases. We can however emphasize that the phase
separation involving the FF regions for the $S_z$ direction will be
easier to be observed with {\it in situ} imaging methods \cite{SIBMIT}
than the correlation functions. 

One of authors (M.O.) wishes to thank R.~Igarashi, N.~Nakai,
H.~Nakamura, and Y.~Ota for fruitful discussion. The work was
partially supported by Grant-in-Aid for Scientific Research
(20500044) from MEXT, Japan. 


\begin{references}
\bibitem{Unote}
The on-site interaction parameter is given by $U = 4 \sqrt{\pi}
(a/a_{\rm lat}) E_{\rm r} [(4 V_{\rm s} + V_{\rm l}) / E_{\rm
r}]^{1/4} (4 V_{\rm s} / E_{\rm r})^{1/4} (V_z / E_{\rm r})^{1/4}$ ($V_0
\gg E_{\rm r}$) for the periodic OL potential $V({\bm r}) = V_{\rm s}
\sin^2 (2kx) + V_{\rm l} \sin^2 (kx) + V_{\rm s} \sin^2 (2ky) + V_z
\sin^2 (kz)$, where $a_{\rm lat}$ is the lattice constant of the
long-period potential, $V_{\rm s}$, $V_{\rm l}$, and $V_{\rm z}$ the
depths of the short-period, long-period and $z$-direction potentials,
respectively, and  $E_{\rm r} = 4 \pi^2 \hbar^2 / 2m a_{\rm lat}^2$ 
the recoil energy \cite{CAreview}.
\end{references}
\end{document}